# Raman Spectroscopy Study of Annealing-Induced Effects on Graphene Prepared by Micromechanical Exfoliation

Jieun Song, Taeg Yeoung Ko, and Sunmin Ryu*

*Department of Applied Chemistry, Kyung Hee University, Yongin, Gyeonggi 446-701, Korea. *E-mail: sunryu@khu.ac.kr*

Since its isolation from graphite,[1] graphene has drawn a lot of experimental and theoretical research. These efforts have been mostly in pursuit of various applications such as electronics,[2] sensors,[3] stretchable transparent electrodes,[4] and various composite materials.[5] To accomplish such graphene-based applications, understanding chemical interactions of this new material with environments during various processing treatments will become more important.[6-9] Since thermal annealing is widely used in various research of graphene for varying purposes such as cleaning,[10, 11] nanostructuring,[12] reactions,[6] etc., understanding annealing-induced effects is prerequisite to many fundamental studies of graphene. In this regard, it is to be noted that there has been a controversy on the cause of the annealing-induced hole doping in graphene.[6, 7, 13] One trivial-looking issue which has not been addressed is the presence of adhesive residues nearby graphene flakes on substrates (see Supporting Information). Although wafer scale graphene samples can now be prepared by chemical vapor deposition methods,[4, 14] the micromechanical exfoliation of graphite using adhesive tape, so called "Scotch tape method",[1] has still been the most reliable and widely used method to generate top-quality graphene samples used for various fundamental studies on graphene. Typical adhesive tape consists of acrylic polymers and various additives detailed composition of which is not revealed in public. When heated or treated with organic solvents, the polymeric residues are likely to contaminate nearby graphene affecting various measurements being taken. So far there appeared many studies which involved either thermal or solvent treatments. For instance, there are several scanning tunneling microscopy studies carried on thermally annealed graphene samples.[10, 11] Device fabrication exploiting e-beam or photo-lithography typically involves solvent treatment and heating in the ambient conditions.[1] There are also studies on chemical reactions of graphene where adhesive residues are likely to interfere with measurements.[1, 9, 15] Thus, annealing-induced effects on graphene prepared by the exfoliation method deserve a careful and systematic study.

In this note, we report a Raman spectroscopy study of annealing-induced effects on graphene samples prepared by the microexfoliation method. It was shown that randomly located adhesive residues often contaminate nearby graphene sheets during thermal annealing. The contamination on graphene can be as thin as ~1 nm, but gives several new Raman bands of unusually strong intensity. We also find that their intensity is strongly dependent on the excitation wavelength implying that graphene-induced Raman enhancement may be operative. The current study also suggests that graphene can be selectively sensitive towards certain molecular species in binding, which can be exploited for interesting application.

Single- and few-layer graphene samples were prepared by the micromechanical exfoliation successfully used in the first isolation of graphene from graphite.[1] Many thin graphite flakes were exfoliated onto adhesive tape by folding and unfolding the tape attached with a piece of high quality kish graphite crystal (Covalent Materials). Freshly cleaved surfaces of graphite flakes located on the tape were then attached to SiO$_2$/Si substrates thoroughly cleaned by piranha solution. The tape was removed from the substrate following gentle rubbing for isolation of graphene. Typically, a few atomically thin graphene flakes were transferred onto the substrate and their size ranged from several to tens of microns across. Owing to the optical interference at the multiple interfaces of the sample and the substrate,[16] even single layer graphene can be clearly located and differentiated from multilayer graphene under an optical microscope. The number of layers (thickness hereafter) of single and few layer graphene samples were determined by Raman spectroscopy as will be discussed below.[17] Raman spectra were obtained in the ambient conditions with a home-built micro Raman setup. Excitation laser beam (633 nm or 514 nm) was focused by a 40x objective onto ~1 micron spot on the sample and backscattered light was collected by the same objective and directed to a spectrograph coupled to a liquid nitrogen-cooled CCD detector. Graphene samples were annealed in a quartz tube furnace under Ar gas flow of 0.70 L/min for two hours at 300 °C.

Figure 1 shows an atomic force microscopy (AFM) image of graphene annealed at 300 °C. The surface of the annealed graphene is very flat except a few foreign particulates marked by horizontal arrows and ~0.5 nm tall elevated regions marked by vertical arrows. In the upper part of the image, there are adhesive residues which are visible under an optical microscope (see Supporting Information). Typically, adhesive residues are not completely removed by annealing in an inert atmosphere. The upper line profile shows that the residues can be as thick as 20 nm. Considering that the residues are in the vicinity of the graphene sheet and the high annealing temperature, it is conceivable that molten polymeric materials may flow and contaminate graphene during annealing process. It is well known that polymeric e-beam resist materials have strong affinity to graphitic materials.[10, 11] In fact, we find that the annealed graphene is unusually thick, 1.8 nm from the lower line profile. However, the thickness of our pristine graphene samples was found to be ~0.7 nm (see Supporting Information) and is consistent with previous measurements which falls within 0.4 ~ 1.0 nm.[1, 11] The increased thickness, ~1 nm corresponding to a few molecular layers, is attributed to binding of foreign materials originating from adhesive residues, which will also be supported by the following Raman scattering study.

Figure 2 shows Raman spectra taken for the graphene sample shown in Fig. 1 before and after the annealing. The intensity of G band at ~1580 cm$^{-1}$ which corresponds to C-C stretching vibration increases almost linearly as increasing thickness. Due to the double resonance mechanism, the line shape of 2D band is closely related to the electronic structure of graphene; the narrow and symmetric 2D Raman band is characteristic of single layer graphene.[17] Following annealing, the 2D-to-G intensity ratio ($I_{2D}/I_G$) decreases from 1.9 to 1.4, while the G and 2D bands upshift by 8 and 12 cm$^{-1}$, respectively. The decreasing $I_{2D}/I_G$ ratio and concurrent upshifts of the two bands can be well explained by hole doping caused by molecular oxygen in the ambient.[18, 19] In addition, it is to be noted that the annealed graphene shows several new bands at < 1550 cm$^{-1}$. The band at 1528 cm$^{-1}$ is even stronger than the G or 2D band of graphene itself and has a very narrow linewidth of 9 cm$^{-1}$. The newly observed Raman bands can be attributed to the materials bound on the graphene as the AFM analysis suggested.

We also note that the new Raman bands are very sensitive to the excitation wavelength; In Fig. 3, three spectra were taken from one identical spot of the annealed graphene shown in Fig. 1, consecutively with 633 (*a*), 514 (*b*), and 633 nm (*c*) excitation lasers. The spectrum (*b*) taken with 514 nm laser does not show any of the new bands but the G band at 1588 cm$^{-1}$. Comparing *a* and *c* spectra, it can be seen that the intensities of the new bands decreased by half due to the irradiation of 514 nm laser used in obtaining *b* spectrum. Repeated measurements with 633 nm laser without irradiation with 514 nm laser showed only minor decrease (<10%) in intensities. Since the employed power density was much larger for 633 nm (5 mW/μm$^2$) than for 514 nm (0.4 mW/μm$^2$), the observed intensity decrease caused by 514 nm excitation is attributed to photochemical degradation of the unidentified materials rather than photothermal effects.

Figure 4 shows the Raman spectrum taken from thick adhesive residues located near the annealed graphene in Fig. 1 (see also Supporting Information). Despite the large thickness of >20 nm, the spectrum contained only very weak bands and thus is shown with the intensity multiplied by a factor of 15. The spectrum consists of a broad band not resolved due to low intensity (centered at 1300 cm$^{-1}$), a band at 1450 cm$^{-1}$, and a multiplet between 2900 and 3000 cm$^{-1}$. While the latter can be assigned to C-H stretching vibrations, the rest cannot be precisely identified based on the given information. More importantly, however, it is to be noted that the adhesive's Raman spectrum is very different from that of the annealed graphene shown in dotted line. While the 1450 cm$^{-1}$ bands in both spectra coincide with each other, the one from the adhesive is much broader (line width ~ 35 cm$^{-1}$).

The above puzzling results may be due to selective Raman enhancement caused by binding with graphene as recently observed for a few different molecular films on graphene.[20] Then the new Raman bands may originate from minor adhesive components of which Raman scattering cross section increases when bound on graphene owing to so called chemical enhancement.[20] Considering the multiplication factor of 15 and the thickness ratio of ~20 between the adhesive residues and the unidentified materials on the annealed graphene, the enhancement factor would be at least 300. The Raman band at 1340 cm$^{-1}$ or 1527 cm$^{-1}$ would give a larger enhancement factor which is known to be affected by the geometry of molecules on surfaces.[20] Another possibility to consider is formation of superlattice on graphene. In this scenario, the unidentified layers on the annealed graphene may form domains of superlattices with larger period than that of graphene. Such superlattices may redefine the Brillouin zone of graphene by so called zone folding. As a result of this, additional phonon modes folded back onto the Brillouin zone center can be detected in a first-order Raman spectrum which allows only for zone-center phonon modes. Some graphite intercalation compounds possess domains of well-ordered in-plane superlattices of intercalants.[21, 22] While detailed analysis of possible superlattices is beyond the scope of the current study, it is to be noted that graphite intercalated with potassium and mercury showed new Raman bands stronger than the G band as seen in Fig. 2.[21]

In summary, Raman spectroscopy was combined with AFM to investigate the effects of thermal annealing on the graphene samples prepared by the widely used micromechanical exfoliation method. Following annealing cycles, adhesive residues were shown to contaminate graphene sheets with thin molecular layers in their close vicinity causing several new intense Raman bands. Detailed investigation shows that the Raman scattering is very strong and may be enhanced by the interaction with graphene. Although the current study does not pinpoint detailed origins for the new Raman bands, the presented results stress that graphene prepared by the above method may require extra cautions when treated with heat or possibly solvents.

**Acknowledgments.** This research was supported by Basic Science Research Program through the National Research Foundation of Korea (NRF) funded by the Ministry of Education, Science and Technology (2009-0089030). We thank Dr. Jihye Shim for insightful comments.

**References**

(1) Novoselov, K. S.; Geim, A. K.; Morozov, S. V.; Jiang, D.; Zhang, Y.; Dubonos, S. V.; Grigorieva, I. V.; Firsov, A. A. *Science* **2004,** *306*, 666-9.
(2) Geim, A. K.; Novoselov, K. S. *Nat. Mater.* **2007,** *6*, 183-191.
(3) Schedin, F.; Geim, A. K.; Morozov, S. V.; Hill, E. W.; Blake, P.; Katsnelson, M. I.; Novoselov, K. S. *Nat. Mater.* **2007,** *6*, 652-655.
(4) Kim, K. S.; Zhao, Y.; Jang, H.; Lee, S. Y.; Kim, J. M.; Kim, K. S.; Ahn, J.-H.; Kim, P.; Choi, J.-Y.; Hong, B. H. *Nature* **2009,** *457*, 706-710.
(5) Wakabayashi, K.; Pierre, C.; Dikin, D. A.; Ruoff, R. S.; Ramanathan, T.; Brinson, L. C.; Torkelson, J. M. *Macromolecules* **2008,** *41*, 1905-1908.
(6) Liu, L.; Ryu, S.; Tomasik, M. R.; Stolyarova, E.; Jung, N.; Hybertsen, M. S.; Steigerwald, M. L.; Brus, L. E.; Flynn, G. W. *Nano Lett.* **2008,** *8*, 1965.
(7) Ryu, S.; Han, M. Y.; Maultzsch, J.; Heinz, T. F.; Kim, P.; Steigerwald, M. L.; Brus, L. E. *Nano Lett.* **2008,** *8*, 4597-4602.
(8) Berciaud, S.; Ryu, S.; Brus, L. E.; Heinz, T. F. *Nano Lett.* **2009,** *9*, 346-352.
(9) Liu, H.; Ryu, S.; Chen, Z.; Steigerwald, M. L.; Nuckolls, C.; Brus, L. E. *J. Am. Chem. Soc.* **2009,** *131*, 17099–17101.
(10) Stolyarova, E.; Rim, K. T.; Ryu, S.; Maultzsch, J.; Kim, P.; Brus, L. E.; Heinz, T. F.; Hybertsen, M. S.; Flynn, G. W. *Proc. Natl. Acad. Sci. U.S.A.* **2007,** *104*, 9209-9212.
(11) Ishigami, M.; Chen, J. H.; Cullen, W. G.; Fuhrer, M. S.; Williams, E. D. *Nano Lett.* **2007,** *7*, 1643-1648.
(12) Bao, W.; Miao, F.; Chen, Z.; Zhang, H.; Jang, W.; Dames, C.; Lau, C. N. *Nat. Nanotechnol.* **2009,** *4*, 562.
(13) Chen, C.-C.; Bao, W.; Theiss, J.; Dames, C.; Lau, C. N.; Cronin, S. B. *Nano Lett.* **2009,** *9*, 4172-4176.
(14) Li, X. S.; Cai, W. W.; An, J. H.; Kim, S.; Nah, J.; Yang, D. X.; Piner, R.; Velamakanni, A.; Jung, I.; Tutuc, E.; Banerjee, S. K.; Colombo, L.; Ruoff, R. S. *Science* **2009,** *324*, 1312-1314.
(15) Sharma, R.; Baik, J. H.; Perera, C. J.; Strano, M. S. *Nano Lett.* **2010,** *10*, 398-405.
(16) Roddaro, S.; Pingue, P.; Piazza, V.; Pellegrini, V.; Beltram, F. *Nano Lett.* **2007,** *7*, 2707-2710.
(17) Ferrari, A. C.; Meyer, J. C.; Scardaci, V.; Casiraghi, C.; Lazzeri, M.; Mauri, F.; Piscanec, S.; Jiang, D.; Novoselov, K. S.; Roth, S.; Geim, A. K. *Phys. Rev. Lett.* **2006,** *97*, 187401/1-187401/4.
(18) Yan, J.; Zhang, Y.; Kim, P.; Pinczuk, A. *Phys. Rev. Lett.* **2007,** *98*, 166802/1-166802/4.
(19) Pisana, S.; Lazzeri, M.; Casiraghi, C.; Novoselov, K. S.; Geim, A. K.; Ferrari, A. C.; Mauri, F. *Nat. Mater.* **2007,** *6*, 198-201.
(20) Ling, X.; Xie, L. M.; Fang, Y.; Xu, H.; Zhang, H. L.; Kong, J.; Dresselhaus, M. S.; Zhang, J.; Liu, Z. F. *Nano Lett.* **2010,** *10*, 553-561.
(21) Timp, G.; Elman, B. S.; Al-Jishi, R.; Dresselhaus, G. *Solid State Commun.* **1982,** *44*, 987-991.
(22) Kelty, S. P.; Lu, Z.; Lieber, C. M. *Phys. Rev. B* **1991,** *44*, 4064.

**Figure captions**

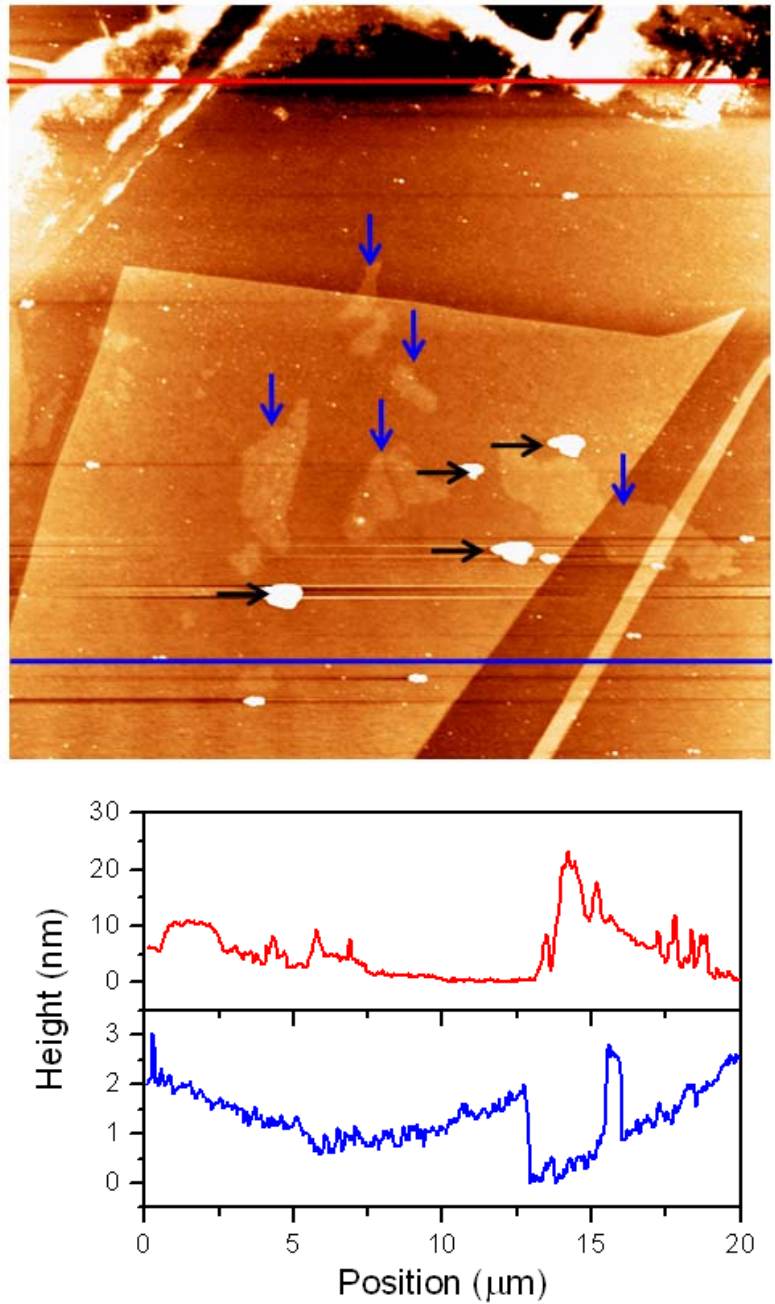

Fig. 1. (a) AFM image of single layer graphene on SiO2/Si substrate, taken after annealing in Ar atmosphere at 300 °C for 2 hrs. The image size is 20 x 20 μm$^2$. (b) Line profiles taken across graphene (lower line) and adhesive residue (upper line).

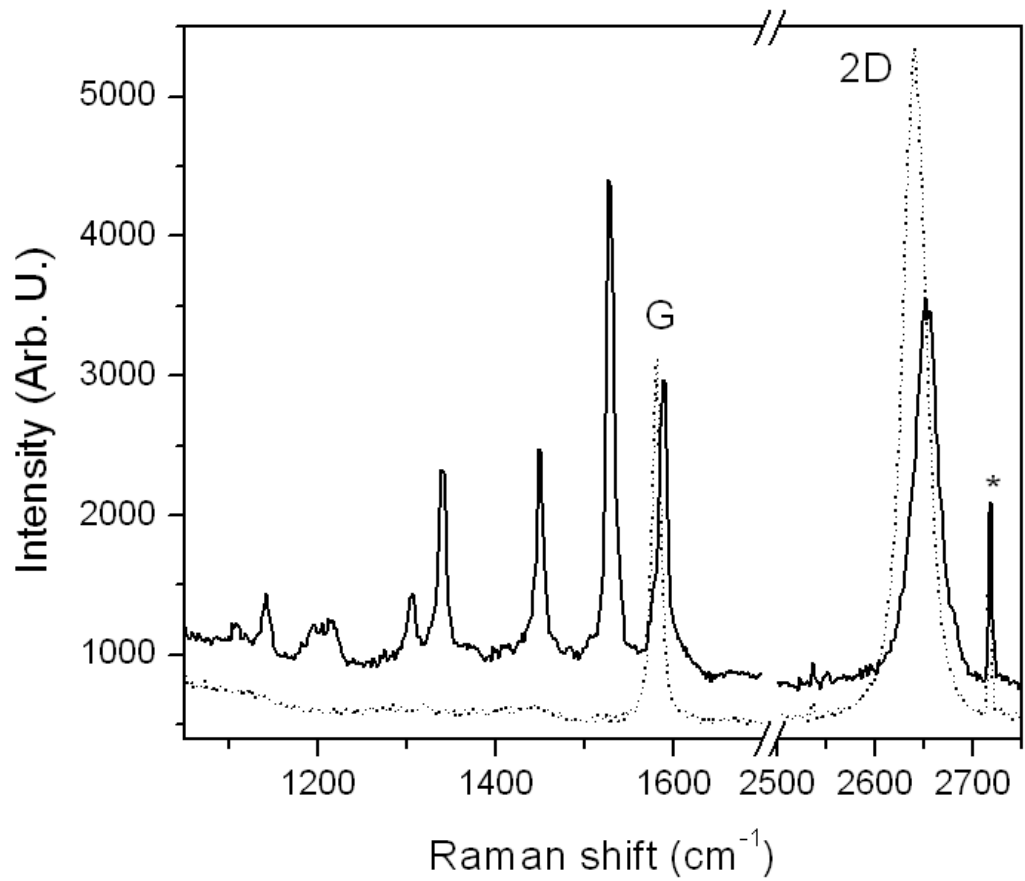

Fig. 2. Raman spectra of the graphene sample shown in Fig. 1, taken before (dotted line) and after (solid line) the thermal annealing. G and 2D denote the Raman G and 2D bands, respectively. The narrow band marked by * originates from a plasma line of the excitation laser operated at 633 nm.

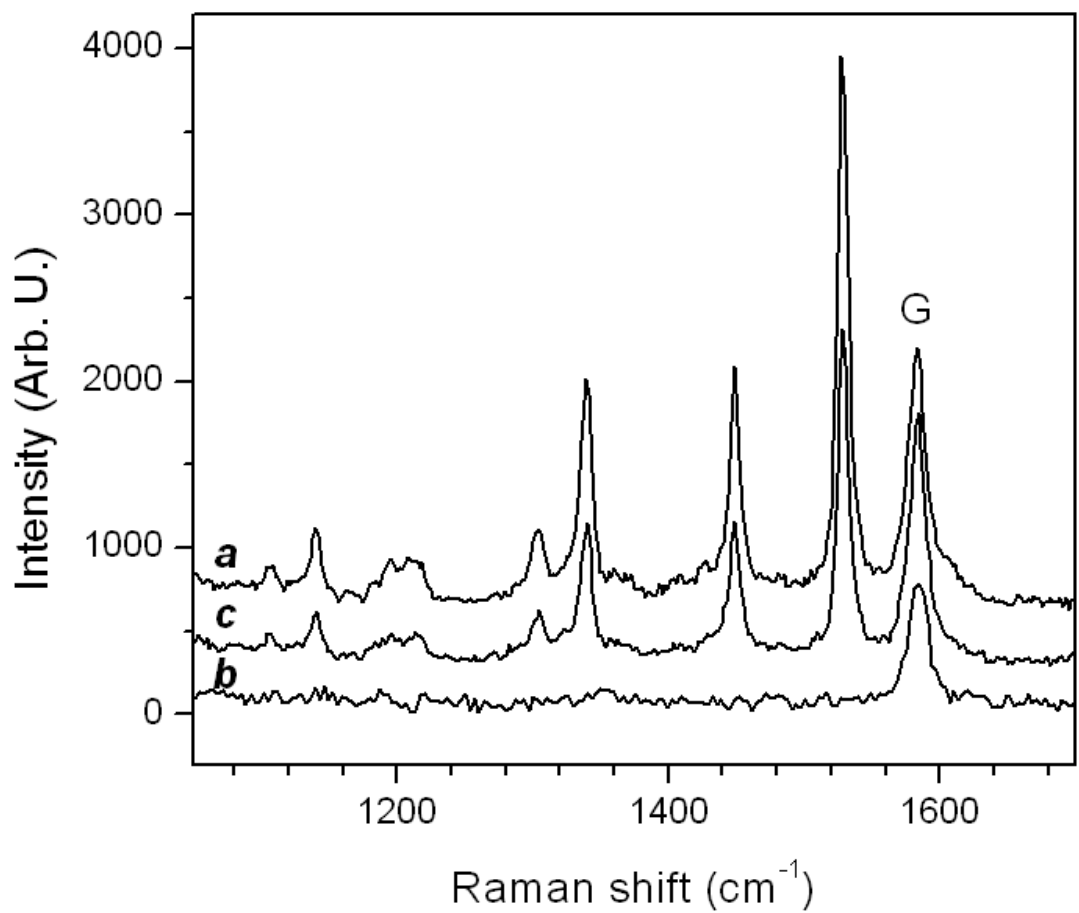

Fig. 3. Raman spectra of the annealed graphene sample shown in Fig. 1, consecutively taken at an identical spot with excitation wavelength of 633 nm (***a***), 514 nm (***b***), and 633 nm (***c***). The annealing-induced bands were not detected with 514 nm excitation and were decreased in intensity by the 514 nm measurement. The spectra were vertically displaced for clarity.

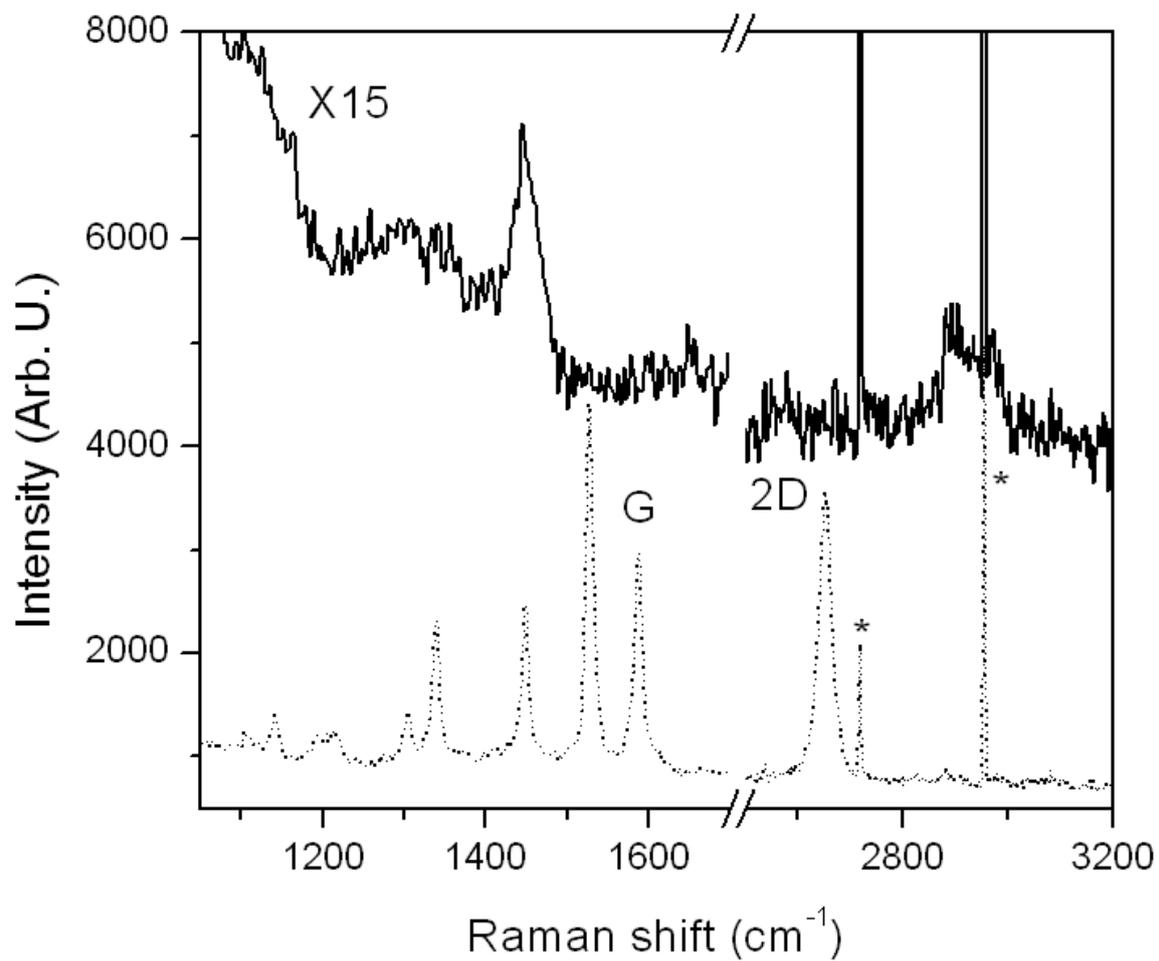

Fig. 4. Raman spectrum (solid line) taken from thick adhesive residue located ~50 μm from the annealed graphene sheet shown in Fig. 1. For comparison, the Raman spectrum of the annealed graphene from Fig. 2 is shown together in dotted line. The narrow bands marked by * originate from plasma lines of the excitation laser operated at 633 nm.